\documentclass[pra,aps,10pt,twocolumn,showpacs]{revtex4-1}

\usepackage{graphicx}
\usepackage{amsmath,amsfonts,bm}

\begin{document}

\title{Semiclassical quantization of the diamagnetic hydrogen atom with 
near action-degenerate periodic-orbit bunches}

\author{Jan Gehrke}
\author{J\"org Main}
\author{G\"unter Wunner}
\affiliation{Institut f\"ur Theoretische Physik 1, Universit\"at Stuttgart,
  70550 Stuttgart, Germany}
\date{\today}

\begin{abstract}
The existence of periodic orbit bunches is proven for the diamagnetic Kepler 
problem.  Members of each bunch are reconnected differently at self-encounters
in phase space but have nearly equal classical action and stability parameters.
Orbits can be grouped already on the level of the symbolic dynamics by 
application of appropriate reconnection rules to the symbolic code in the 
ternary alphabet.  The periodic orbit bunches can significantly improve the 
efficiency of semiclassical quantization methods for classically chaotic 
systems, which suffer from the exponential proliferation of orbits.
For the diamagnetic hydrogen atom the use of one or few representatives 
of a periodic orbit bunch in Gutzwiller's trace formula allows for the
computation of semiclassical spectra with a classical data set reduced by
up to a factor of 20.
\end{abstract}


\maketitle

\section{Introduction}
\label{sec:intro}
Semiclassical theories provide the link between quantum spectra and the
dynamics of the underlying classical system.
For systems with chaotic classical dynamics Gutzwiller's trace formula
\cite{Gut90} expresses the density of states in terms of parameters of the 
isolated unstable periodic orbits.
Chaotic systems typically exhibit an exponential proliferation of the
number of periodic orbits with growing length.
The periodic-orbit sum of Gutzwiller's trace formula usually does not 
converge, and special techniques as, for example, the cycle expansion 
\cite{Cvi89} or harmonic inversion \cite{Mai97c,Mai98b,Mai99d} 
method must be applied for the computation of semiclassical eigenvalues
from a finite set of trajectories.
Although these techniques can be very powerful for certain systems 
a remaining conceptually weak point of periodic orbit quantization is 
still that, due to the exponential proliferation of orbits, the gradual
improvement or extension of semiclassical spectra usually requires an
exponentially increasing set of periodic orbit data.
In this article we propose a way to significantly reduce the number 
of trajectories which is necessary for periodic orbit quantization 
by employing properties of the orbits which have recently been 
investigated in connection with universal features of quantum chaos.

According to the Bohigas-Giannoni-Schmit conjecture \cite{Boh84} 
a signature of quantum chaos is universal spectral fluctuations on 
the scale of the mean level spacing.
For example, for systems with time reversal symmetry the spacings show 
a Wigner distribution.
On the way towards an understanding or even proof of the conjecture
the correlations between orbits which play an important role in the
semiclassical approximation to the spectral form factor have been 
studied  \cite{Sie01,Sie02,Mue04,Heu04,Mue05,Heu07,Mue09}.
Strong correlations exist only between orbits which have identical or
near identical actions.
It has been found that periodic orbits consist of different segments 
\cite{Alt08}.
In each segment, an orbit follows closely its neighboring orbit or the
time-reverse of this orbit, but the orbits differ in how these segments
are connected at self-encounters.
As a consequence long periodic orbits of hyperbolic systems do not exist
as independent individuals but rather come in closely packed
near action-degenerate periodic-orbit bunches.
Pairs of orbits with two differently connected loops provide the leading
off-diagonal contributions to the form factor \cite{Sie01,Sie02}.
The role of self-encounters and periodic-orbit bunches for universal
level correlations in quantum chaos has been investigated in detail in
Refs.\ \cite{Mue04,Heu04,Mue05,Heu07,Mue09}.

Here we want to demonstrate that near action-degenerate periodic-orbit 
bunches can help to significantly improve the efficiency of semiclassical
quantization methods.
The idea is not to use all individual periodic orbits up to a given length
but only one or very few representatives of a periodic-orbit bunch.
Those representatives are appropriately weighted in the periodic orbit sum
according to the size of the bunch, which can be determined, for systems
with a known symbolic code, solely on the level of the symbolic dynamics
without a numerically expensive periodic-orbit search.

Results are presented for the hydrogen atom in a magnetic field, which 
has a long history as a prototype model of a real physical system with 
signatures of quantum chaos \cite{Has89,Fri89}.
At sufficiently high positive energies it is an open system with fully
hyperbolic classical dynamics, and a unique symbolic code for the
periodic orbits does exist \cite{Eck90,Han95,Schnei97}.
Semiclassical resonances of the diamagnetic hydrogen atom have already been 
obtained with a modified and extended cycle expansion technique \cite{Tan96}.

The article is organized as follows.
In Sec.~\ref{sec:Hamilton} we discuss the classical dynamics of the 
diamagnetic hydrogen atom and present an efficient multi-shooting algorithm
for the numerical periodic-orbit search.
In Sec.~\ref{sec:com_rules} we reveal the existence of periodic-orbit bunches
and introduce four reconnection rules which allow for the grouping of orbits
on the level of the symbolic dynamics.
Semiclassical resonances are computed in Sec.~\ref{sec:harm_inv} with the
harmonic inversion method, and it is illustrated how using periodic-orbit 
bunches significantly improves the efficiency of the method.
Concluding remarks are given in Sec.~\ref{sec:conclusion}.

\section{Classical dynamics and periodic-orbit search}
\label{sec:Hamilton}
In this section we recapitulate the basic equations for the classical dynamics
of the diamagnetic hydrogen atom and present a very efficient multi-shooting
algorithm for the periodic orbit search.

In atomic units the Hamiltonian of the diamagnetic hydrogen atom reads
\begin{equation}
 H = \frac{1}{2}{\bm p}^2 - \frac{1}{r} + \frac{1}{2}\gamma L_z
   + \frac{1}{8}\gamma^2(x^2+y^2) = E \; ,
\label{AG-1}
\end{equation}
where $\gamma=B/(2.35\times 10^5{\rm T})$ with $B$ the magnetic field strength.
We only consider states with magnetic quantum number $m=L_z=0$.
The classical dynamics does not depend separately on the energy $E$
and the magnetic field strength $\gamma$ but only on the scaled energy
$\tilde E=E\gamma^{-2/3}$.
Using a regularization of the Coulomb singularity in scaled semiparabolical 
coordinates \cite{Has89,Fri89}
\begin{equation}
 \mu = \gamma^{1/3} \sqrt{r+z} \text{   and   }
 \nu = \gamma^{1/3} \sqrt{r-z} \; , 
\label{AG-5}
\end{equation}
the classical equations of motion in the transformed time $\tau$, with
$d\tau=2r dt$, are derived from the scaled and regularized Hamiltonian
\begin{equation}
 h = \frac{1}{2}\left(p_{\mu}^2+p_{\nu}^2\right) + V(\mu,\nu) = 2 \; ,
\label{AG-6}
\end{equation} 
with the potential
\begin{equation}
 V(\mu,\nu) = - \tilde E\left(\mu^2+\nu^2\right) 
   + \frac{1}{8}\left(\mu^4\nu^2+\mu^2\nu^4\right) \; . 
\label{eq:pot}
\end{equation} 
The numerical computation of periodic orbits is a prerequisite for
the semiclassical quantization of chaotic systems.
In chaotic billiards with an existing symbolic dynamics such as the 
three-disk scatterer or the hyperbola billiard, a periodic orbit with
given symbolic code can be computed efficiently by moving the reflection
points until the orbit length becomes a minimum.
For systems with smooth potentials the periodic orbit search is more difficult.
Using a simple shooting algorithm trajectories with varying starting points
are integrated numerically until the initial and final point match.
When applying this method for a systematic periodic orbit search some
orbits are typically found many times while others may be overlooked.
In particular, long and very unstable orbits are hard to find.

For the computation of the periodic orbits of the diamagnetic hydrogen atom
we employ a multi-shooting algorithm which is adapted to the symbolic
dynamics of the system.
With this algorithm we can find selectively each orbit corresponding
to a given symbolic code, even when the orbit is very long and unstable.

If one allows the semiparabolic coordinates $(\mu,\nu)$ to be positive or 
negative, the potential (\ref{eq:pot}) has a $C_{4v}$ symmetry.
Periodic orbits can be described by a ternary symbolic code \cite{Eck90}
in a similar way as for the hyperbola or four-disk billiard with the same 
symmetry.
At high energies $\tilde E > \tilde E_c=0.329$ there is a one-to-one
correspondence between orbits and the symbolic code, whereas below the
critical energy orbits undergo bifurcations and the symbolic dynamics is
pruned \cite{Han95,Schnei97}.

\begin{figure}
\includegraphics[width=0.75\columnwidth]{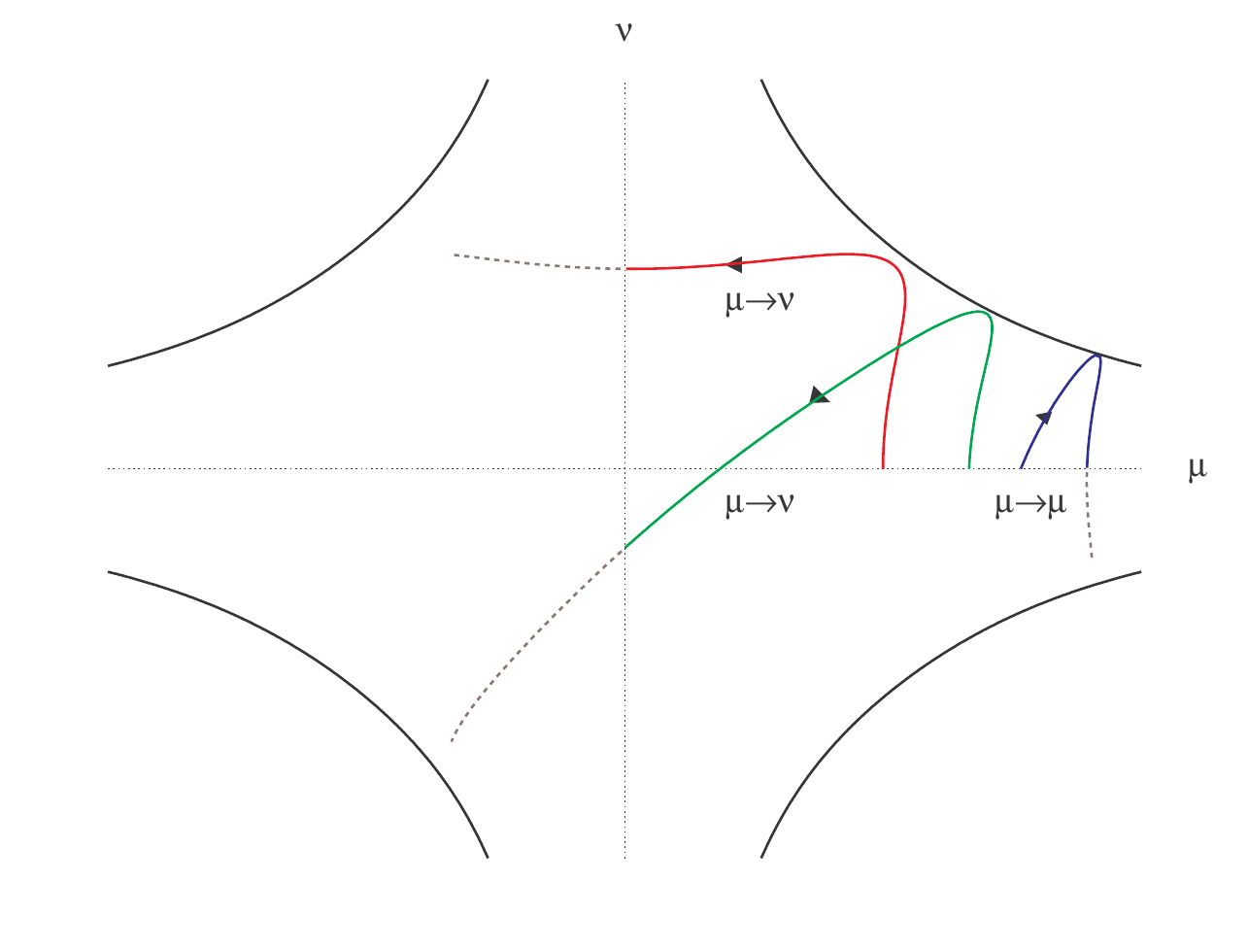}
\caption{(Color online)
 In the multi-shooting algorithm each orbit is split into short segments
 starting and ending on one of the axes.  The concatenation of various types
 of segments is guided by the symbolic code of the orbit.}
\label{fig1}
\end{figure}

The multi-shooting algorithm for the periodic orbit search basically works
as follows.
Each periodic orbit is split into segments, with the number of segments 
equal to the string length $L$ of the symbolic code.
The segments start on either the $\mu$ or the $\nu$ axis and end on one of
the axes as illustrated in Fig.~\ref{fig1}.
When we start all segments with initial guesses chosen in accordance with 
the symbolic code the orbit is discontinuous between the end point 
of one and the starting point of the next segment.
Now the initial conditions of all segments are iteratively changed to
remove the discontinuities.
This problem can be formulated as a $2L$-dimensional root search.
(Note that two parameters can be changed for each segment to vary the
initial conditions on one of the axes.)

Because the segments are short the multi-shooting algorithm usually 
converges rapidly and uniquely to the periodic orbit selected by the
symbolic code, provided the orbit is not pruned or very close to
bifurcation.
The method works very well even for very long and unstable orbits.

\begin{figure*}
\includegraphics[width=0.31\textwidth]{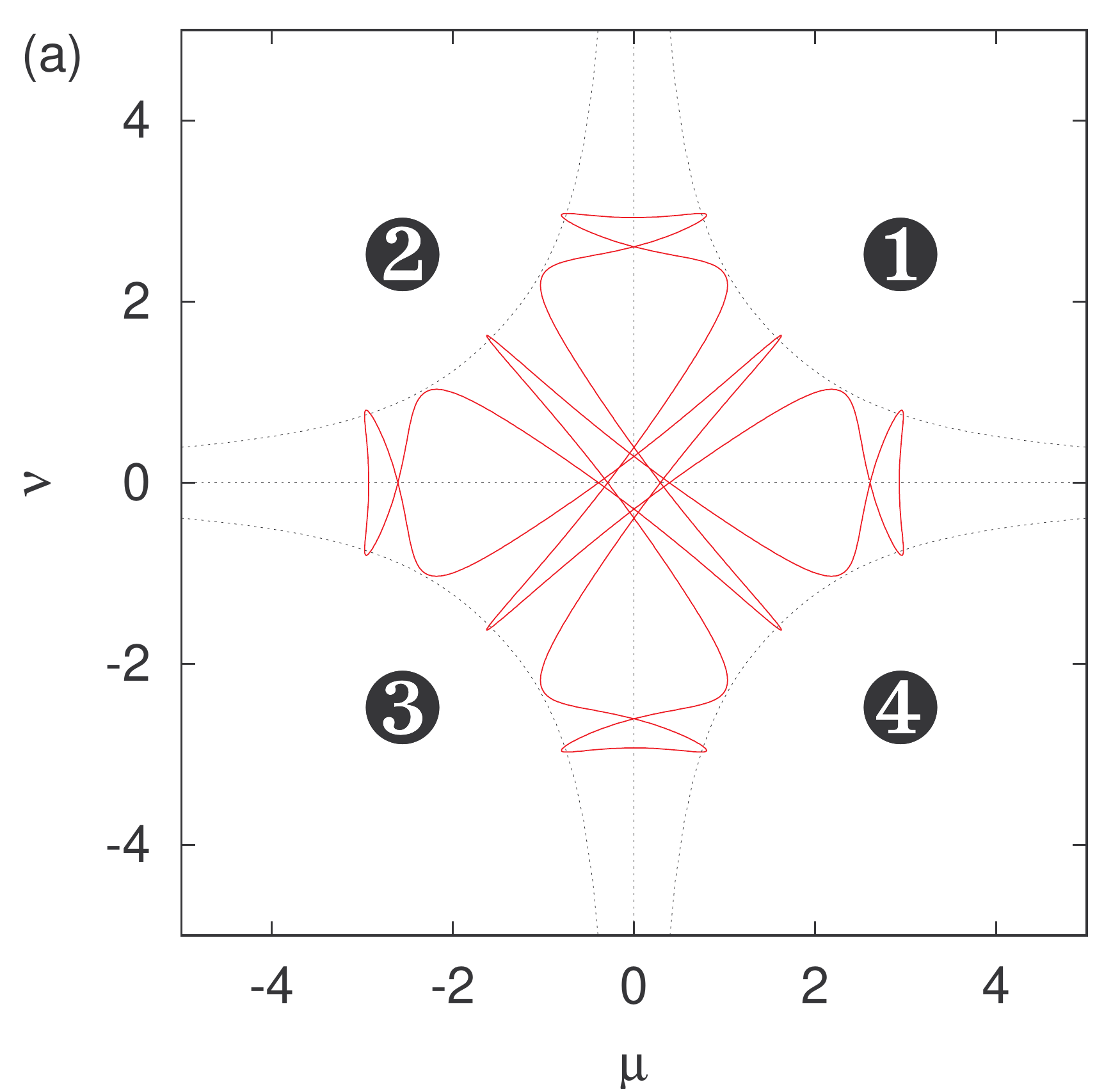}
\includegraphics[width=0.31\textwidth]{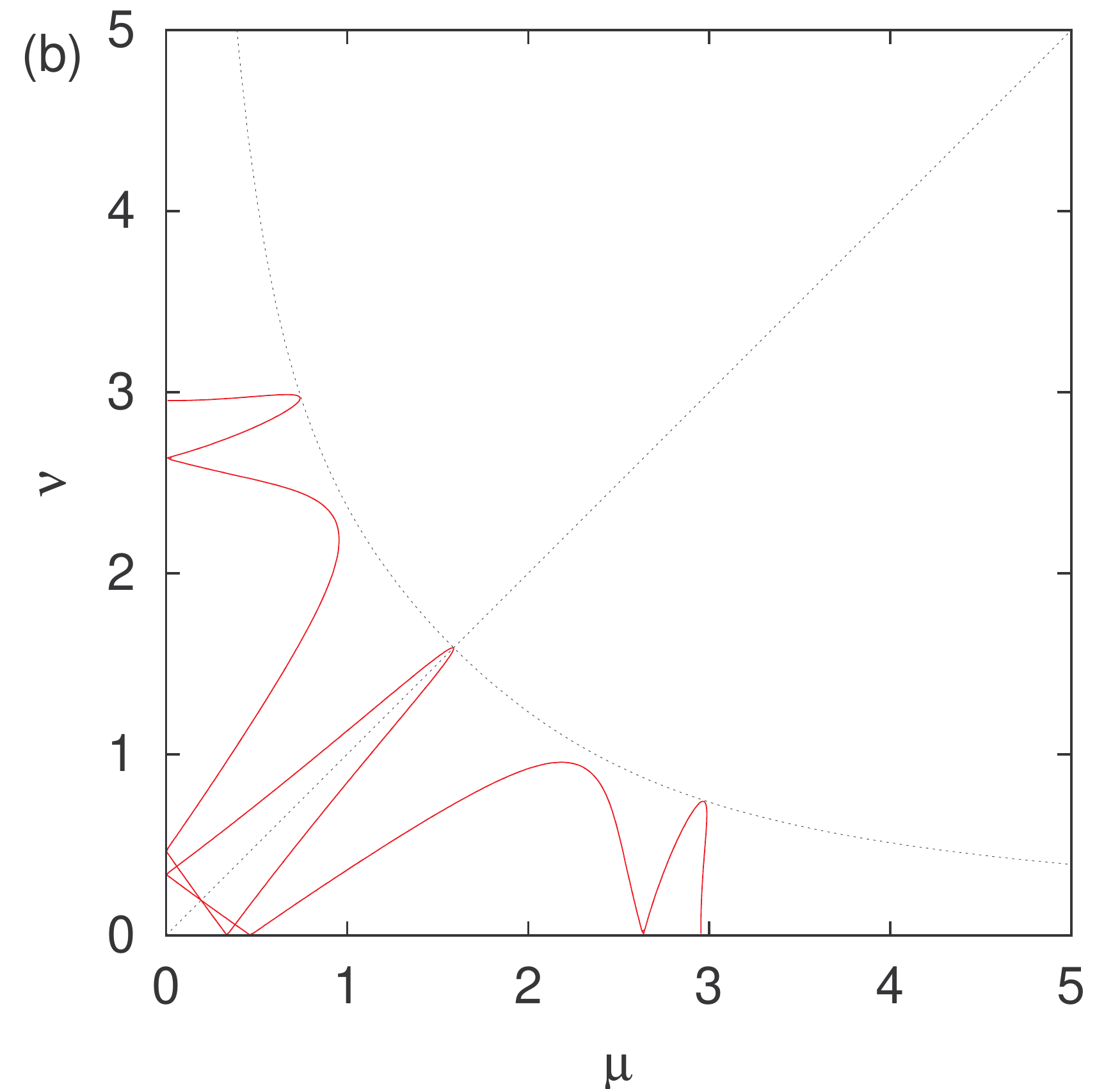}
\includegraphics[width=0.31\textwidth]{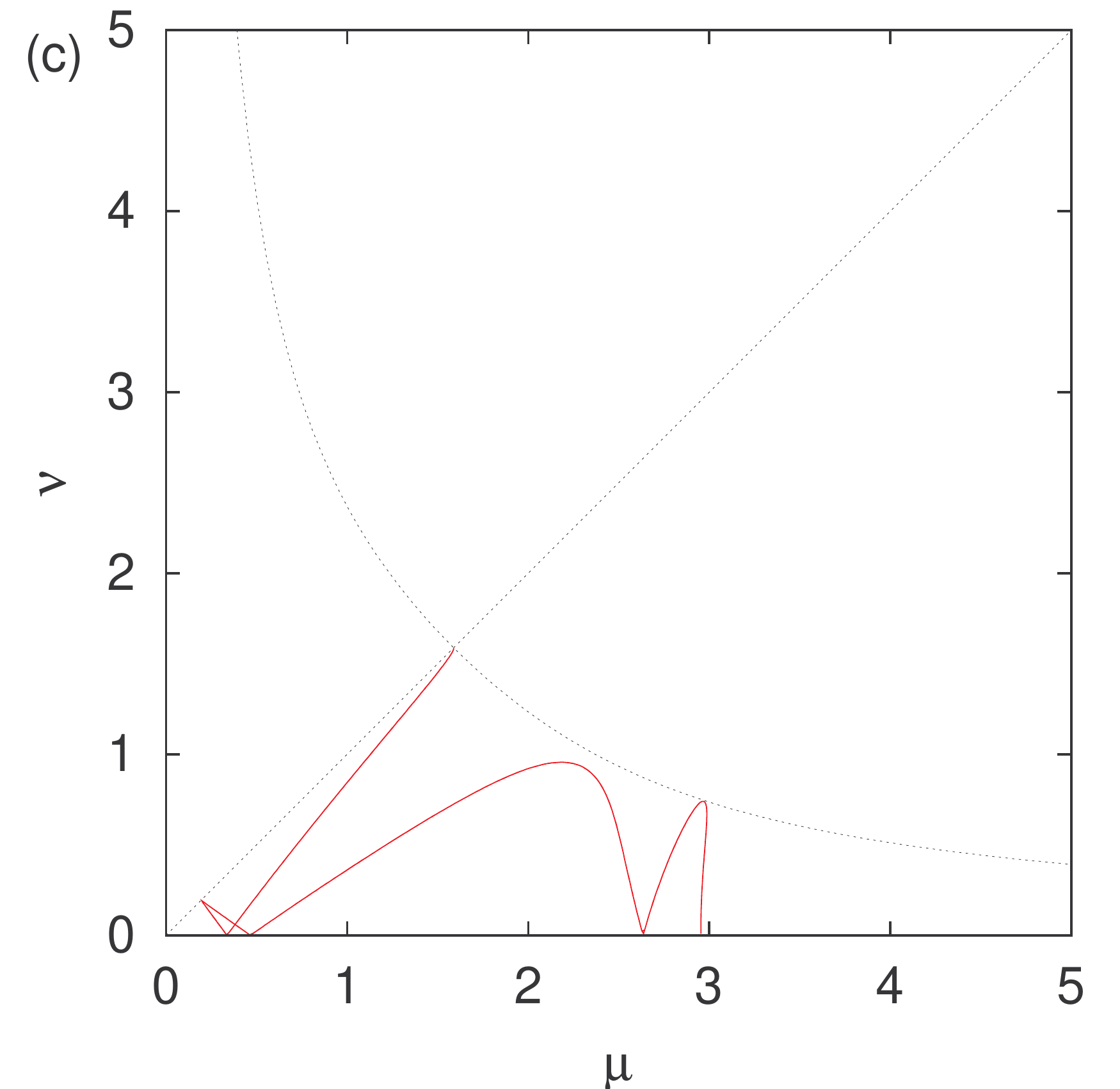}
\caption{(Color online)
 Example of a periodic orbit displayed in (a) the full coordinate plane,
 (b) the 1st quadrant, and (c) the fundamental domain.}
\label{fig2}
\end{figure*}

An example of a periodic orbit at scaled energy $\tilde E=0.5$ is 
presented in Fig.~\ref{fig2}.
For the symbolic code of a periodic orbit we adopt the ternary alphabet
of Ref.~\cite{Eck90} with a slightly modified way of the symmetry reduction.
In the full coordinate space [see Fig.~\ref{fig2}(a)] where the
semiparabolic coordinates are extended to negative values each orbit
is labeled by a sequence of numbers 1 to 4 for the reflections at
the potential barrier in one of the quadrants.
The $C_{4v}$ symmetry of the potential allows for symmetry reduction of
the symbolic code.
Symbols $\mu$, $\nu$, and \texttt{0} are used for the orbit crossing the
$\mu$- or $\nu$-axis in a clockwise or anticlockwise turn, or moving
to the opposite potential barrier with a crossing of both axis.
At this level of symmetry reduction orbits are located in the first
quadrant [see Fig.~\ref{fig2}(b)].
Finally, an axis symbol $\mu$ or $\nu$ is replaced with a \texttt{+} or
\texttt{-} sign when it follows a different or the same axis symbol
(ignoring any intermediate \texttt{0} symbols), respectively.
This alphabet describes the symmetry reduced orbit in the fundamental domain
[see Fig.~\ref{fig2}(c)].
The various levels of symmetry reduction are helpful to illustrate the 
periodic orbit bunches and reconnection rules in Sec.~\ref{sec:com_rules}.

\section{Periodic-orbit bunches}
\label{sec:com_rules}
In this section we first provide an example of a group of near 
action-degenerate orbits of the diamagnetic hydrogen atom.
We then present four reconnection rules which are based on the symbolic 
dynamics of the periodic orbits and can be used for grouping the 
trajectories in periodic-orbit bunches.
The properties of the periodic-orbit bunches are discussed.

\subsection{Example of  a near action-degenerate  periodic-orbit bunch}
\label{subsec:example}
\begin{figure}
\includegraphics[width=0.9\columnwidth]{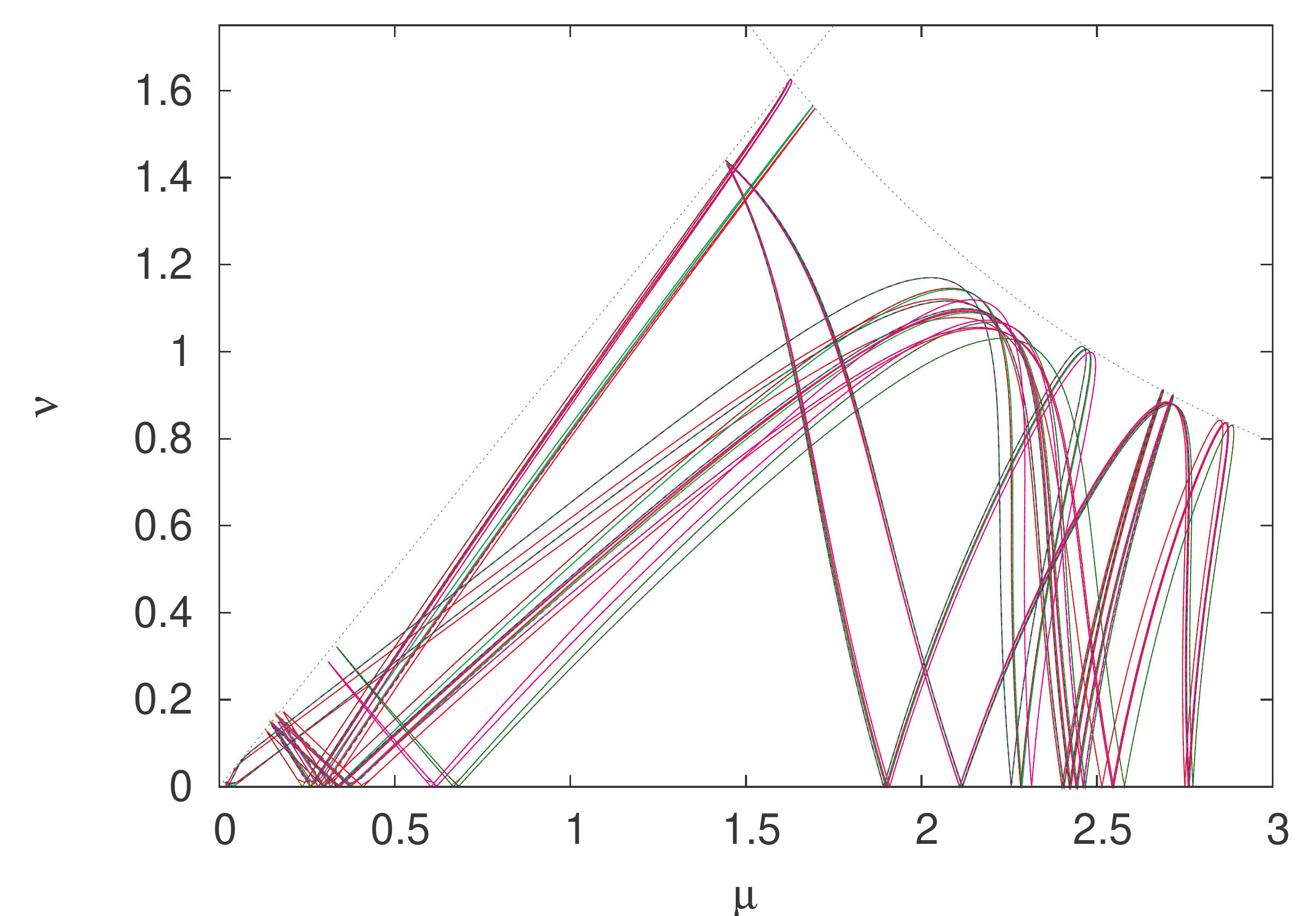}
\caption{(Color online)
 Example of a periodic-orbit bunch consisting of 16 near action-degenerate
 orbits drawn in the fundamental domain of the $(\mu, \nu)$ coordinate space
 for scaled energy $\tilde E=0.5$. The similarities between orbits are
 obvious in the overview.}
\label{fig3}
\end{figure}

We start our discussion of periodic-orbit bunches of the diamagnetic hydrogen
atom by way of example of a group of 16 orbits with cycle length $L=10$
at scaled energy $\tilde E=0.5$.
The orbits are presented in Fig.\ \ref{fig3} in the fundamental domain of
the coordinate space.
In the overview the similarities between the orbits are quite obvious.
All orbits appear to run nearly parallel and thus to be located in the 
same area of the phase space.
\begin{figure*}
\includegraphics[width=0.97\textwidth]{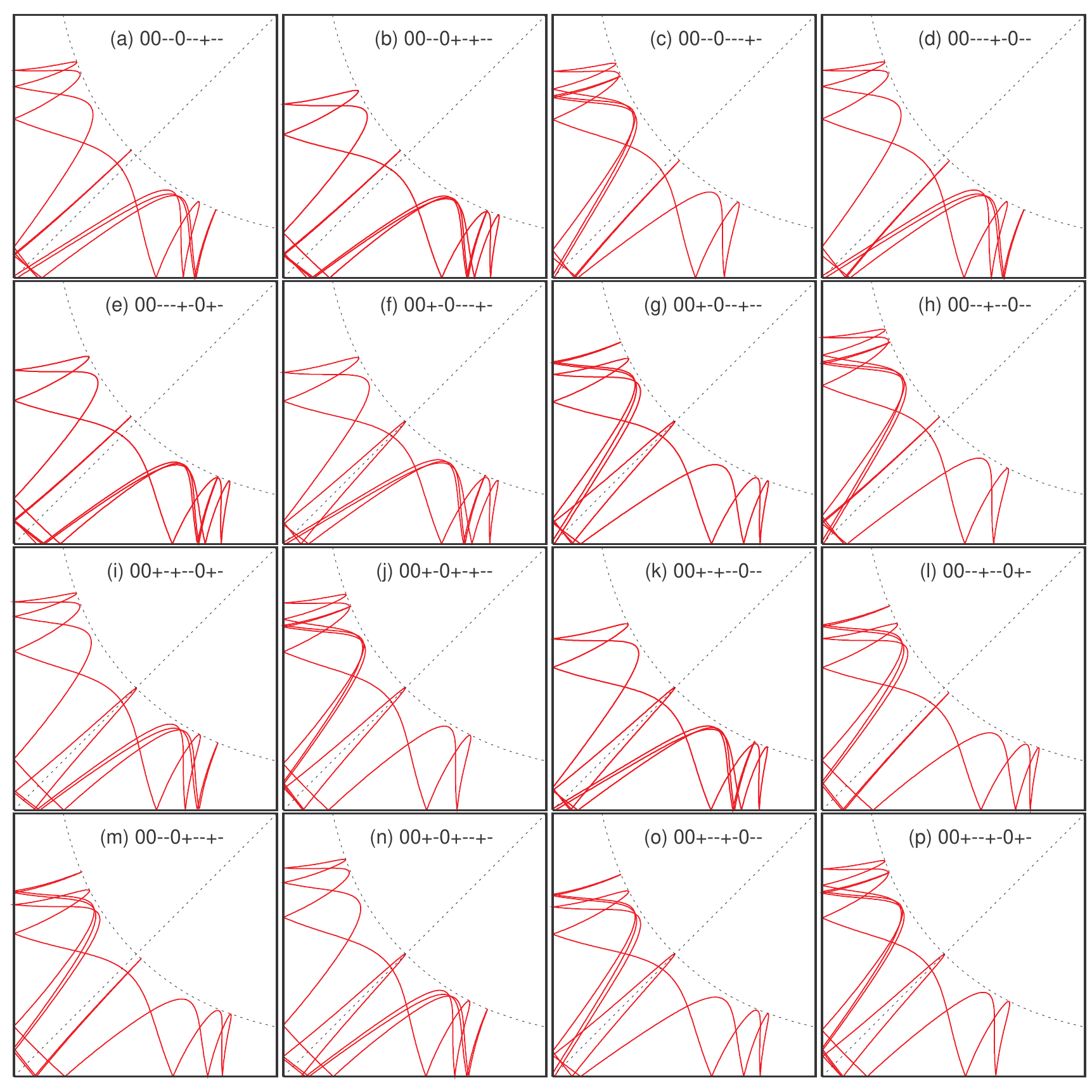}
\caption{(Color online)
 Example of a near action-degenerate periodic-orbit bunch at scaled energy
 $\tilde E=0.5$ consisting of 16 orbits with cycle length $L=10$.  The
 trajectories are drawn in semiparabolic $(\mu, \nu)$ coordinates and are
 labeled by the symbolic code.  See text for discussion.}
\label{fig4}
\end{figure*}
For a more detailed discussion the 16 orbits of the periodic-orbit bunch
are drawn separately in Fig.~\ref{fig4} in semiparabolic $(\mu, \nu)$ 
coordinates without symmetry reduction along the $\mu=\nu$ symmetry line.
Each orbit is labeled by the symbolic code using the ternary alphabet
introduced in Sec.\ \ref{sec:Hamilton}.
To make the graphs concise only one fundamental period is shown in 
Fig.~\ref{fig4}, i.e., some orbits are periodic only in the fully symmetry
reduced fundamental domain.
At first glance the orbits in Fig.~\ref{fig4} do not appear to be very 
similar, however, the similarities are obvious in Fig.\ \ref{fig3} after 
complete symmetry reduction in the fundamental domain of the coordinate space.
A more detailed inspection of Fig.~\ref{fig4} allows us to reveal
both the similarities and differences of the members of the periodic-orbit 
bunch and to illustrate the reconnection rules.

The basic mechanism for the formation of near action-degenerate periodic-orbit 
bunches is that each periodic orbit can be split into segments where the
starting and end points of segments can be connected in various ways using
simple combinatorial rules.
Two types of segments exist where the trajectory is located either close to
or far away from the $\mu=\nu$ symmetry line.
They can be described most efficiently by the symbolic code of the orbits.
A sequence of one or more zero symbols followed by one plus or minus
symbol characterizes a segment located near the $\mu=\nu$ symmetry line,
a sequence of plus and minus symbols (without any zero) characterizes
a segment away from the symmetry line.
All orbits in  Fig.~\ref{fig4} thus consist of four segments, e.g.,
$\texttt{OO-}|\texttt{-}|\texttt{O-}|\texttt{-+--}$
in Fig.\ \ref{fig4}(a).
The orbits in Fig.\ \ref{fig4}(b)-(p) are obtained from the orbit in 
Fig.~\ref{fig4}(a) by reconnecting the four segments in various ways.
For example, in the orbit 
$\texttt{OO-}|\texttt{-+--}|\texttt{O-}|\texttt{-}$
in Fig.~\ref{fig4}(h) the two segments {\tt -} and {\tt -+--} are interchanged, 
and in the orbit {\tt OO-}$|${\tt -}$|${\tt O-}$|${\tt --+-} in 
Fig.~\ref{fig4}(c) the segment {\tt -+--} is traversed backward.
Furthermore, the plus or minus symbol following a sequence of zero
symbols can be interchanged [see orbits 
$\texttt{OO-}|\texttt{-}|\texttt{O+}|\texttt{-+--}$,
$\texttt{OO+}|\texttt{-}|\texttt{O-}|\texttt{-+--}$, and 
$\texttt{OO+}|\texttt{-}|\texttt{O+}|\texttt{-+--}$ 
in Fig.\ \ref{fig4}(b), (g), and (j)] which is related to the mirroring 
of segments at the $\mu=\nu$ symmetry line.
The remaining orbits in Fig.~\ref{fig4} are obtained by combined application 
of several reconnection rules, e.g., in Fig.~\ref{fig4}(k) the segments
\texttt{-} and \texttt{-+--} are interchanged and the first segment 
\texttt{OO-} is replaced with \texttt{OO+}.

\subsection{Reconnection rules}
\label{subsec:rules}
We denote a sequence of one or more zero symbols followed by one plus
or minus symbol as \texttt{(0)}-stretch, and an arbitrary sequence of 
plus and minus symbols without any zero symbol as {\tt (+-)}-stretch.
The grouping of trajectories of the diamagnetic hydrogen atom in periodic-orbit
bunches can now be described by the following four reconnection rules.

{\em Rule R-1:}
The last symbol of a \texttt{(0)}-stretch, which is a plus or minus symbol, 
can be replaced with the other one.

{\em Rule R-2:}
The order of symbols in a \texttt{(+-)}-stretch can be reversed.

{\em Rule R-3:}
In a symbolic code with two or more \texttt{(0)}-stretches different 
\texttt{(0)}-stretches can be interchanged.

{\em Rule R-4:}
In a symbolic code with two or more \texttt{(+-)}-stretches different 
\texttt{(+-)}-stretches can be interchanged.

To illustrate the reconnection rules we discuss once more the periodic orbits 
shown in Fig.~\ref{fig4}.
The first orbit {\tt 00--0--+--} in Fig.~\ref{fig4}(a) can be decomposed into 
the {\tt (0)}-stretch {\tt 00-}, the {\tt (+-)}-stretch {\tt -}, the 
{\tt (0)}-stretch {\tt 0-}, and the {\tt (+-)}-stretch {\tt -+--}.
The remaining 15 orbits in Fig.~\ref{fig4} can be obtained from that orbit
by application of one or more of the reconnection rules.
For example, the orbit in Fig.~\ref{fig4}(b) is obtained by applying rule
R-1 to the second {\tt (0)}-stretch {\tt 0-}, the orbit in Fig.~\ref{fig4}(c)
is obtained by applying rule R-2 to the {\tt (+-)}-stretch {\tt -+--}, and
the orbit in Fig.~\ref{fig4}(d) is obtained by applying the rules R-2 and
R-3, i.e., the {\tt (+-)}-stretch {\tt -+--} is reversed and then the two
{\tt (+-)}-stretches {\tt --+-} and {\tt -} are interchanged.
Applications of the reconnection rules mean that parts of the orbit
drawn in semiparabolic $(\mu,\nu)$ coordinates are approximately mirrored
at the $\mu=\nu$ symmetry line, parts of the orbit are traversed backward,
or parts of the orbit are visited in a different order.
However, in the fundamental domain of the $(\mu,\nu)$ coordinate space
the global shape of orbits belonging to the same periodic-orbit bunch
is very similar, as can be clearly seen in Fig.~\ref{fig3}.

\begin{figure}
\includegraphics[width=\columnwidth]{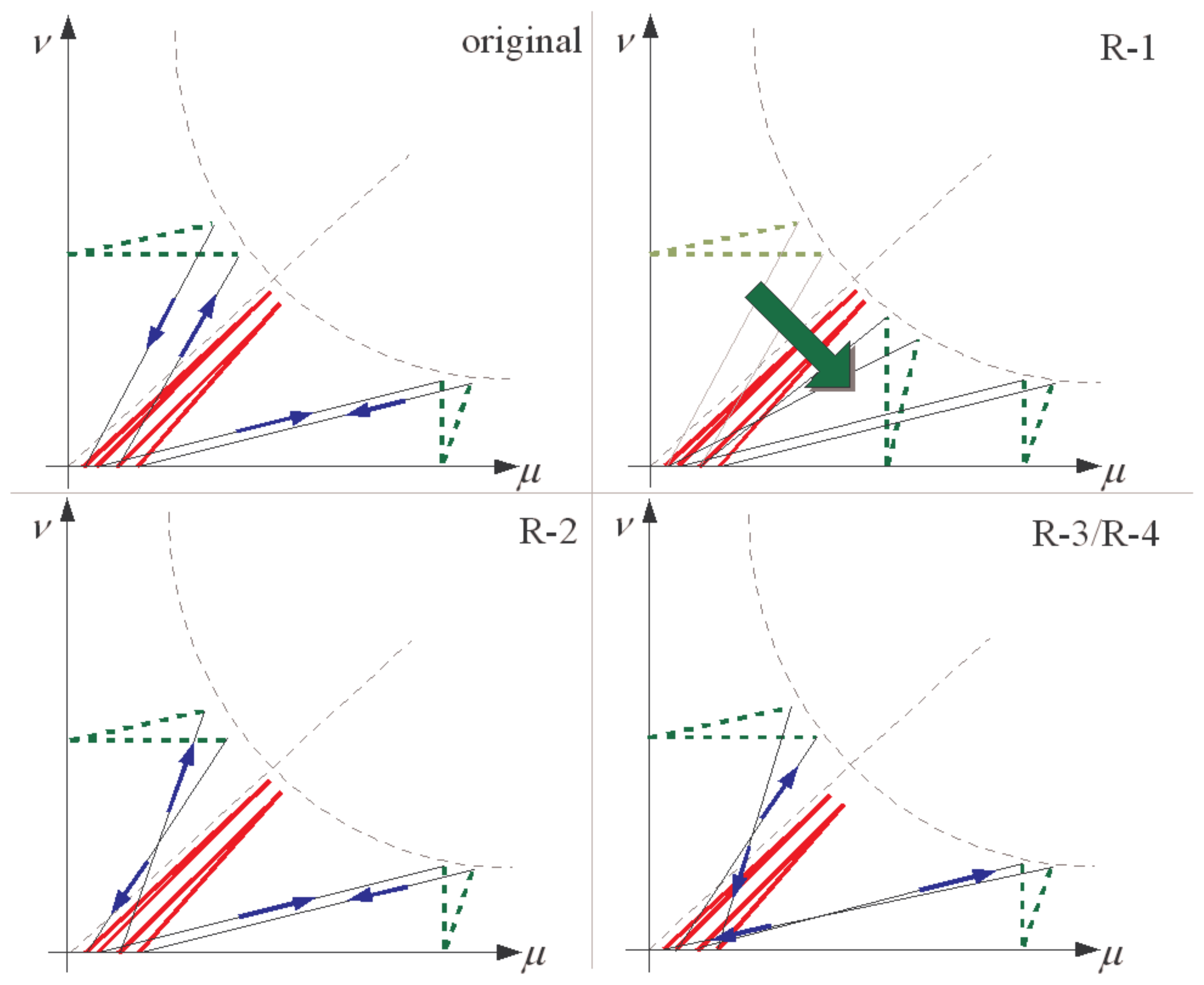}
\caption{(Color online)
 Sketch of the four reconnection rules.  In the $(\mu,\nu)$ coordinate
 space \texttt{(0)}-stretches indicated by red solid lines are located
 in the vicinity of the $\mu=\nu$ symmetry line, and {\tt (+-)}-stretches
 indicated by green dashed lines are located in regions with $\mu\ne\nu$
 away from the symmetry line.  The connections between \texttt{(0)}- and
 \texttt{(+-)}-stretches are drawn with thin black lines.
 \texttt{(+-)}-stretches can be mirrored at the $\mu=\nu$ symmetry
 line (rule R-1), traversed backward (rule R-2), and the orders of
 the \texttt{(0)}- or \texttt{(+-)}-stretches can be interchanged
 (rules R-3 and R-4).}
\label{fig5}
\end{figure}
The reconnection rules are schematically illustrated in Fig.~\ref{fig5}.
A periodic-orbit bunch is characterized by a set of $k$ {\tt (+-)}-stretches,
which is a sequence of plus and minus symbols (an empty set is allowed),
and (for $k\ge 2$) the same number of {\tt (0)}-stretches.
The {\tt (0)}-stretches are located in the vicinity of the $\mu=\nu$ symmetry 
line and are indicated by red solid lines in Fig.~\ref{fig5}.
The {\tt (+-)}-stretches are located in regions with $\mu\ne\nu$ away from 
the symmetry line and are indicated by green dashed lines.
Each {\tt (0)}-stretch is followed by a {\tt (+-)}-stretch and vice versa,
the connections are indicated by the thin black lines in Fig.~\ref{fig5}.
The individual members of the periodic-orbit bunch differ in the way how
the stretches are connected:
{\tt (+-)}-stretches can be mirrored at the $\mu=\nu$ symmetry line (rule R-1),
traversed backward (rule R-2), and the orders of the {\tt (0)}- or 
{\tt (+-)}-stretches can be interchanged (rules R-3 and R-4).

How many orbits belong to a certain periodic-orbit bunch?
Although we cannot give a precise answer it is easy to estimate an upper
bound of that number from the combinatorial reconnection rules.
Let again $k$ be the number of {\tt (+-)}- and {\tt (0)}-stretches in a 
symbolic code.
According to rule R-1 the last symbol of a {\tt (0)}-stretch, which is 
{\tt +} or {\tt -}, can be replaced with the other one, resulting in $2^k$ 
possibilities.
According to rule R-2 the order of symbols in a {\tt (+-)}-stretch can be
reversed.
Each {\tt (+-)}-stretch which is not symmetric under that transform provides
a factor of two, resulting in at most $2^k$ possibilities.
The interchange of {\tt (0)}-stretches and {\tt (+-)}-stretches according
to rules R-3 and R-4 provide at most (substrings may be identical) $k!$ 
and $(k-1)!$ possibilities, respectively, where, in the second factorial, 
it has been taken into account that a cyclic permutation of all {\tt (0)}- 
and {\tt (+-)}-stretches does not change the periodic orbit.
Altogether we obtain a maximum number of orbits 
\begin{equation}
 N_k^{\max} = k!\, (k-1)!\, 2^{2k}
\end{equation}
building one periodic-orbit bunch.
The 16 orbits shown in Fig.~\ref{fig4} have $k=2$ \texttt{(0)}-stretches.
The number of orbits is less than the maximum $N_2^{\max}=32$ because the
\texttt{(+-)}-stretch consisting of a single minus symbol coincides with 
its reverse.
The maximum number $N_k^{\max}$ increases very rapidly with $k$, e.g.,
for $k=4$ bunches with up to $N_4^{\max}=36864$ orbits exist.
However, for such a huge bunch the four \texttt{(0)}-stretches must have 
different lengths and the four \texttt{(+-)}-stretches including their time 
reversals must be different, which is found to be possible only for very 
long orbits with code lengths $L\ge 26$.

\subsection{Properties of periodic-orbit bunches}
\label{subsec:bunch-properties}
\begin{figure}
\includegraphics[width=0.9\columnwidth]{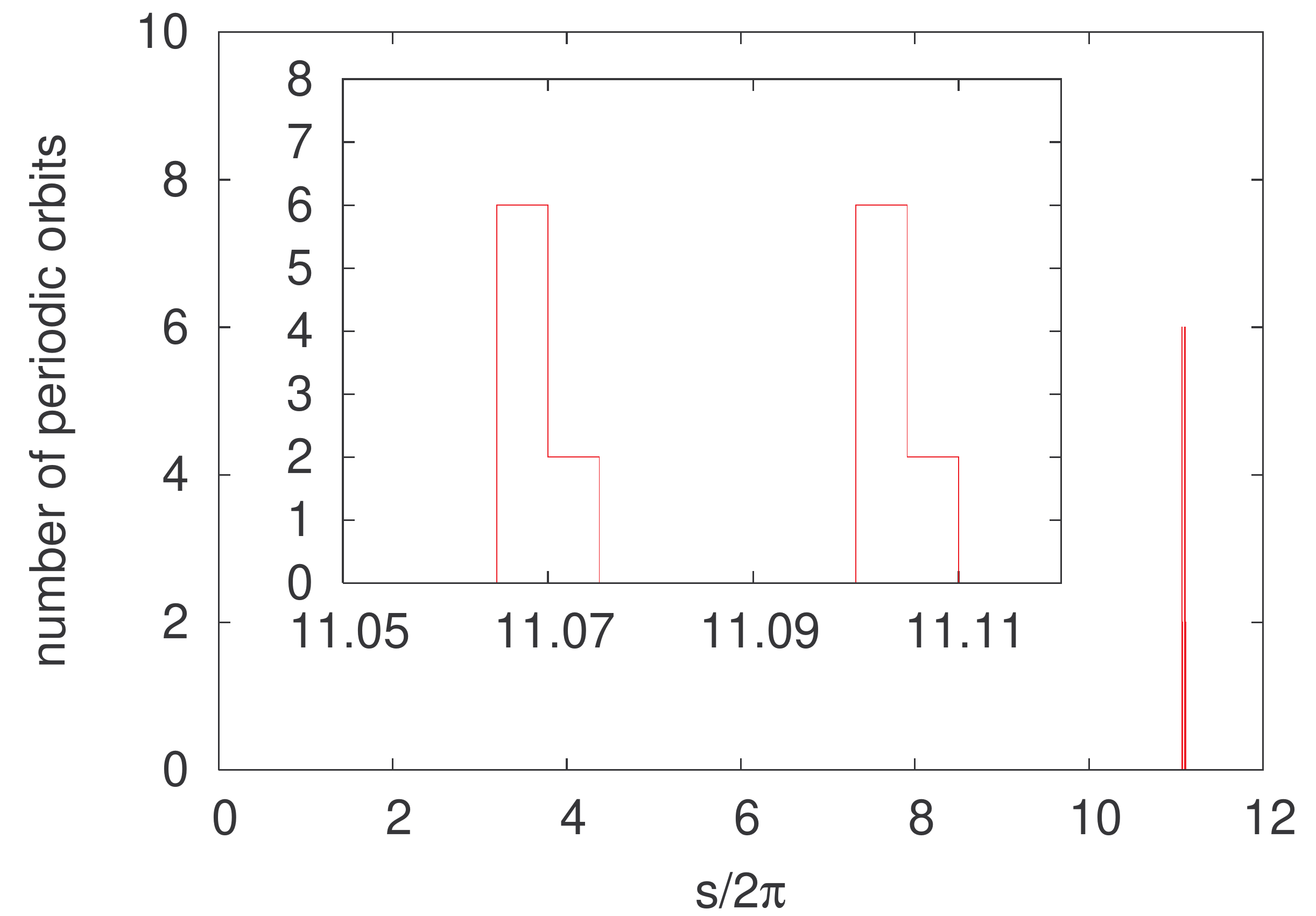}
\caption{(Color online)
 Distribution histogram of the classical actions for the 16 trajectories of
 the near action-degenerate periodic-orbit bunch shown in Figs.~\ref{fig3}
 and \ref{fig4}.  The small differences between the actions of individual
 orbits are only visible in the enlarged inset.}
\label{fig6}
\end{figure}

The most important feature of the periodic-orbit bunches is that the individual
trajectories have similar shape in coordinate space (see Fig.~\ref{fig3}),
and thus all orbits have nearly the same action and stability properties.
For the 16 orbits shown in Figs.~\ref{fig3} and \ref{fig4} the near
degeneracy of the actions is illustrated in Fig.~\ref{fig6}.
In low resolution the distribution of the actions exhibits a single peak
at $s/2\pi\approx 11.1$.
Only in high resolution (see the inset in Fig.~\ref{fig6}) the distribution
of the actions in a small but nonzero range $\Delta s$ can be observed.

\begin{figure}
\includegraphics[width=0.9\columnwidth]{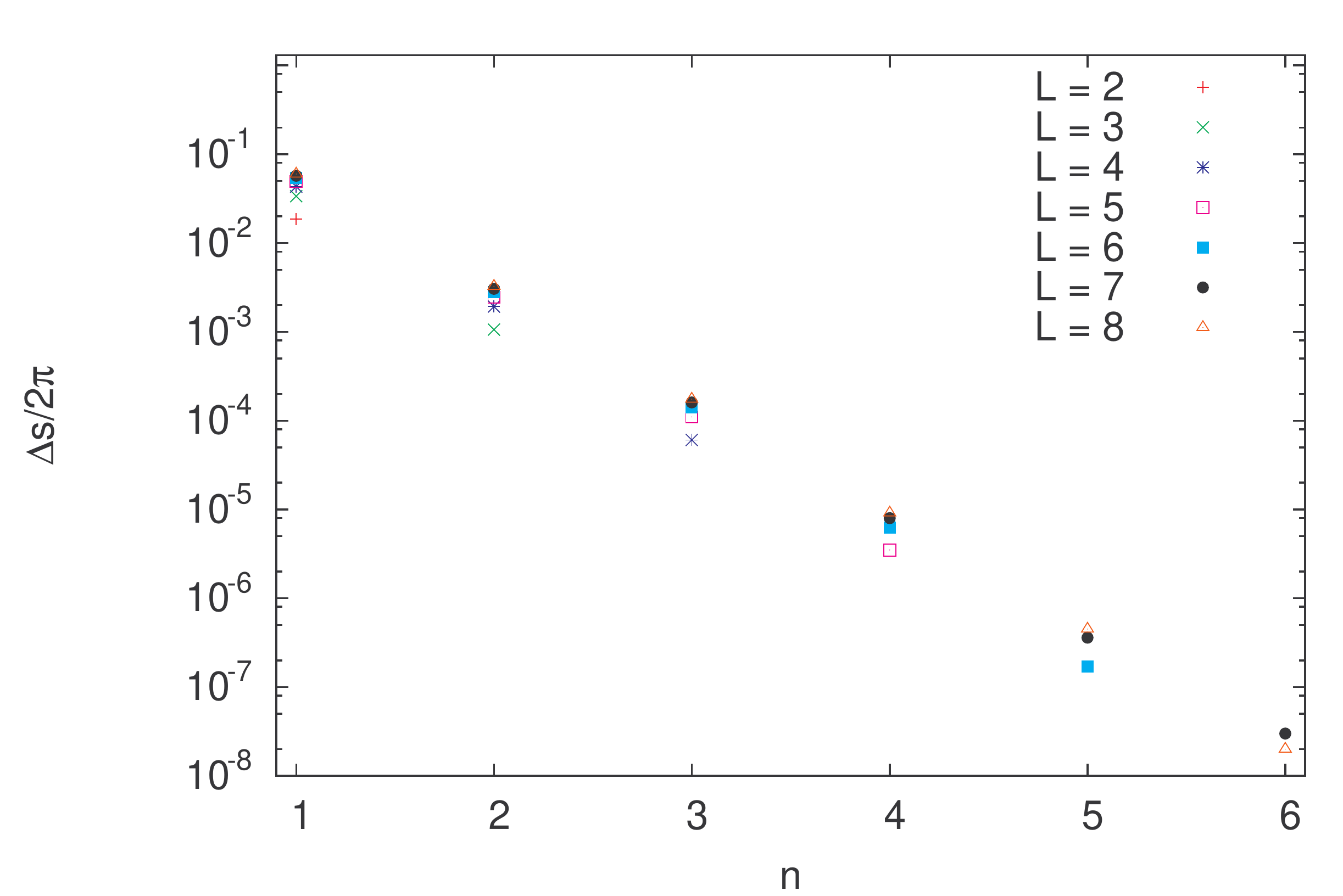}
\caption{(Color online)
 Action difference $\Delta s$ between two orbits consisting of a 
 \texttt{(0)}-stretch with $n$ zero symbols and a \texttt{(+-)}-stretch with 
 $L-n-1$ minus symbols.  Evidently, $\Delta s$ decreases exponentially
 with the length of the \texttt{(0)}-stretch.}
\label{fig7}
\end{figure}

For a given periodic-orbit bunch the action range $\Delta s$ basically
depends on the number of zero symbols in the shortest \texttt{(0)}-stretch.
We illustrate this for the orbit pairs
\begin{align*}
 \underbrace{\texttt{0000}}_{n}{\texttt{-}}\underbrace{\texttt{------}}_{L-n-1}
 \quad \text{and} \quad
 \underbrace{\texttt{0000}}_{n}{\texttt{+}}\underbrace{\texttt{------}}_{L-n-1}
\end{align*}
which consist of a \texttt{(0)}-stretch with $n$ zero symbols followed by a 
plus or minus symbol and a \texttt{(+-)}-stretch with $L-n-1$ minus symbols.
The total length of the symbolic code is $L$.
For fixed $L$ and $n$ the two orbits belong to the same periodic-orbit
bunch (see rule R-1).
The action difference $\Delta s$ between the two orbits for various values
of $L$ and $n$ is shown in Fig.~\ref{fig7}.
Evidently, $\Delta s$ decreases exponentially with the number $n$ of
consecutive zero symbols but only weakly depends on the total code
length $L$.
This result does not depend on the special choice of the \texttt{(+-)}-stretch
as a sequence of minus symbols, i.e., similar results have been obtained 
for other types of \texttt{(+-)}-stretches.
A more detailed investigation shows that the similarities of orbits in a 
periodic-orbit bunch become more and more pronounced with increasing length 
of the shortest \texttt{(0)}-stretch.
The reason is the decrease of the angle or distance of the orbit at 
self-encounters with increasing length of the \texttt{(0)}-stretches 
\cite{Gehrke08}.

\section{Semiclassical quantization with periodic-orbit bunches}
\label{sec:harm_inv}
As already mentioned in Sec.~\ref{sec:intro} self-encounters and periodic-orbit
bunches play an important role for the understanding of the level statistics
of quantum systems with an underlying chaotic classical dynamics.
We will now demonstrate the relevance of periodic-orbit bunches for the
semiclassical quantization of the diamagnetic hydrogen atom.

The energy eigenvalues of integrable classical systems can be obtained by 
semiclassical torus quantization.
In chaotic systems torus structures in phase space are absent.
It has been shown by Gutzwiller \cite{Gut90} that the unstable periodic
orbits are the skeleton for the semiclassical quantization of chaotic systems.
However, the problems with Gutzwiller's trace formula are twofold:
First, the periodic orbit sum is not absolutely convergent and special
techniques such as cycle expansion \cite{Cvi89} 
or harmonic inversion \cite{Mai97c,Mai98b,Mai99d}
methods are necessary to obtain converged results.
Second, in chaotic systems the number of periodic orbits grows exponentially
with the orbital lengths.
It is very unsatisfactory that an exponentially increasing set of classical 
data information is necessary to resolve a few more semiclassical eigenvalues.
However, as already mentioned, the required classical data can be 
significantly reduced when using the periodic-orbit bunches instead of 
the individual periodic orbits for the semiclassical quantization.

\subsection{Harmonic inversion method}
\label{subsec:harm}
We employ the harmonic inversion method for periodic orbit quantization
\cite{Mai97c,Mai98b,Mai99d,Mai00}, and, for the convenience of the reader, 
here briefly recapitulate its basic ingredients.
According to \cite{Gut90} the semiclassical response function of chaotic 
systems is given by
\begin{align}
 g_{\text{sc}}(E) = g_{\text{sc}}^0(E)
 + \sum_{\text{po}}\mathcal{A}_{\text{po}}(E)\, e^{iS_{\text{po}}(E)} \; , 
\label{trace_formula}
\end{align}
where $g_{\text{sc}}^0$ is a smooth function of the energy,
$S_{\text{po}}$ is the classical action of a periodic orbit, and 
$\mathcal{A}_{\text{po}}$ is the amplitude of that orbit including phase 
information.
The semiclassical energies or resonances are the poles of the response 
function $g_{\text{sc}}(E)$ in Eq.~\eqref{trace_formula}.

The hydrogen atom in a magnetic field possesses a scaling property, as already
introduced in Sec.~\ref{sec:Hamilton}, i.e., the classical dynamics does not 
depend separately on the energy $E$ and the magnetic field strength $\gamma$ 
but only on the scaled energy $\tilde E=E\gamma^{-2/3}$.
Keeping the scaled energy constant the response function can be written as
a function of $w=\gamma^{-1/3}$, viz.
\begin{align}
 g_{\text{sc}}(w) = g_{\text{sc}}^0(w)
 + \sum_{\text{po}}\mathcal{A}_{\text{po}}\, e^{iws_{\text{po}}} \; , 
\label{trace_formula_scaled}
\end{align}
where $s_{\text{po}}=\gamma^{1/3}S_{\text{po}}$ is the scaled action of a periodic
orbit and the amplitude
\begin{align}
 \mathcal{A}_{\text{po}} = \frac{s_{\text{ppo}}}
 {\sqrt{ | 2 -\lambda_{\text{ppo}}^r - \lambda_{\text{ppo}}^{-r} | }}\,
 e^{-ir\mu_{\text{ppo}}\pi/2}
\label{loading}
\end{align}
depends on the scaled action $s_{\text{ppo}}$ of the primitive periodic orbit
(ppo) where the orbit is traversed only once, the repetition number $r$ of 
that orbit, the leading eigenvalue $\lambda_{\text{ppo}}$ of the monodromy 
matrix, and the Maslov index $\mu_{\text{ppo}}$.
The Fourier transform of the scaled response function 
in Eq.~\eqref{trace_formula_scaled} yields the semiclassical signal
\begin{align}
 C_{\text{sc}}(s)=\frac{1}{2\pi}\int_{-\infty}^{\infty}g_{\text{sc}}(w) e^{-is{w}}d w
 = \sum_{\text{po}}\mathcal{A}_{\text{po}}\delta(s-s_{\text{po}}) , 
\label{propagator}
\end{align} 
which possesses $\delta$-peaks with weight factors $\mathcal{A}_{\text{po}}$ 
at the actions $s=s_{\text{po}}$ of the classical periodic orbits.

The periodic orbit quantization is achieved by adjusting the semiclassical
signal \eqref{propagator} to its quantum analogue.
The quantum resonances are the eigenvalues $w=\gamma^{-1/3}$ of 
the scaled Schr\"odinger equation at constant scaled energy
\begin{align}
  &\left[-2\tilde E\left(\mu^2+\nu^2\right)
  + \frac{1}{4}\left(\mu^4\nu^2+\mu^2\nu^4\right) -4\right] \psi \nonumber \\
  &= w^{-2} \left(\Delta_\mu +\Delta_\nu\right) \psi \; \text{ with } \;
  \Delta_\rho=\frac{1}{\rho}\frac{\partial}{\partial\rho}
  \left(\rho\frac{\partial}{\partial\rho}\right) \; .
\label{eq:gen_ev}
\end{align}
The Fourier transform of the quantum mechanical Green's function,
\begin{align}
 g_{\text{qm}}(w) = \sum_k \frac{d_k}{w-w_k} \; , 
\label{G_6-4}
\end{align}
with resonances in the lower complex half plane, i.e., ${\rm Im}\,w_k<0$ and
residuals $d_k=1$ for non-degenerate states yields the quantum signal
\begin{align}
 C_{\text{qm}}(s)=\dfrac{1}{2\pi}\int_{-\infty}^{\infty}g_{\text{qm}}(w)e^{-isw}dw
 = -i\sum_k d_k e^{-iw_k s} \; .
\label{qm-versions}
\end{align}
The semiclassical signal \eqref{propagator} obtained from the classical
periodic orbit data is now adapted to the quantum signal \eqref{qm-versions},
with the $d_k$ and $w_k$ being free adjustable complex parameters by means of
nonlinear signal processing methods.
Technical details can be found in Refs.\ \cite{Mai00,Belkic00a}.

The periodic orbit signal $C_{\text{sc}}(s)$ must be known in the interval
$0<s<s_{\max}$.
The required signal length $s_{\max}$ is proportional to the density of
eigenstates $w_k$, which usually increases for higher excitations, i.e.,
an increasing signal length is required to resolve the denser parts of 
the quantum spectrum.

\subsection{Reduced data set}
\label{subsec:dataset}
\begin{figure}
\includegraphics[width=0.9\columnwidth]{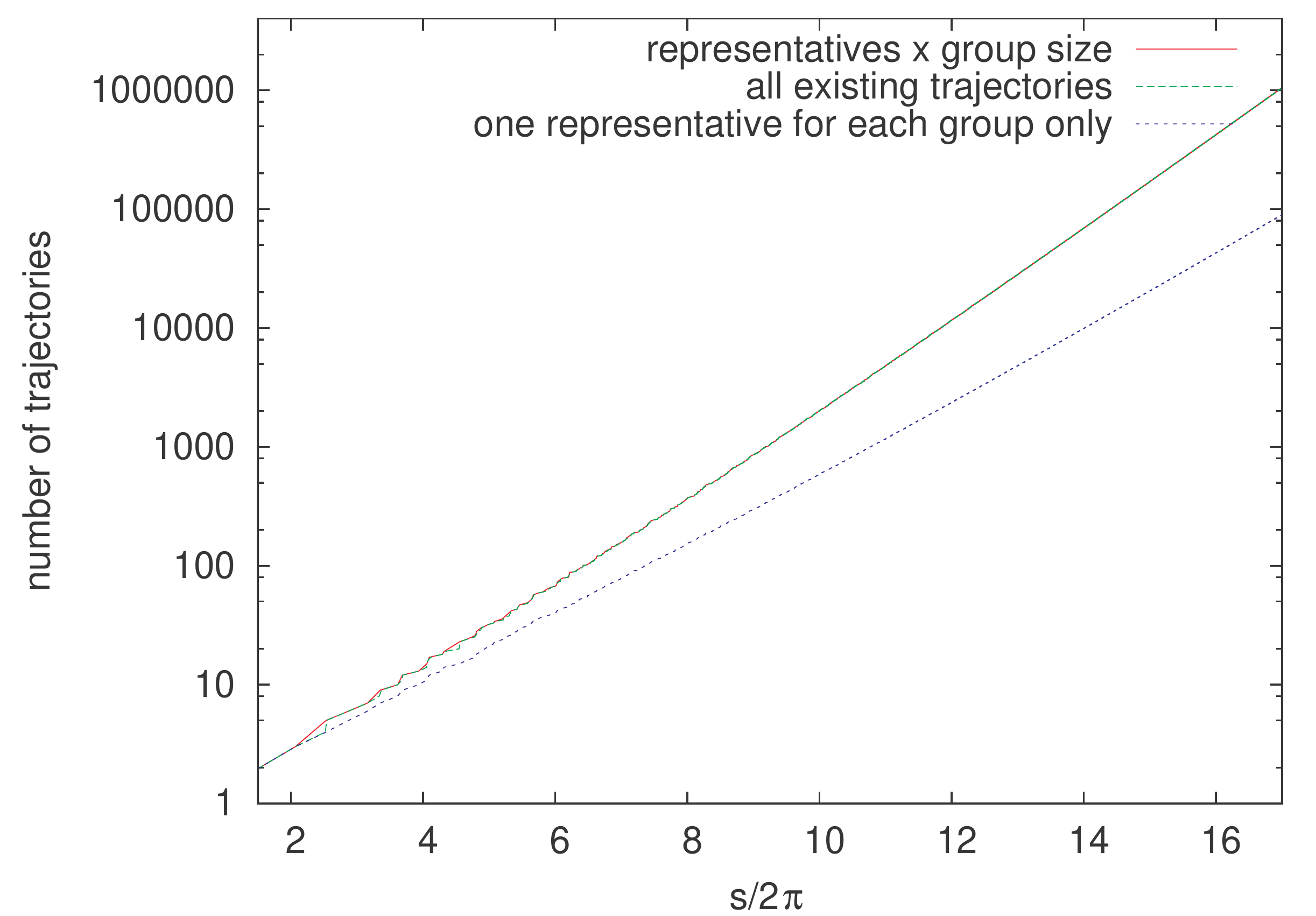}
\caption{(Color online)
 Number of periodic orbits 
 with classical action $s_{\text{po}}<s$ for all individual orbits
 (dashed green line), only one representative of a periodic-orbit bunch
 (dotted blue line), and one representative  of a periodic-orbit bunch
 weighted with the size of the bunch (solid red line, nearly indistinguishable
 from the dashed green line).  All curves show an exponential increase,
 however, the number of periodic-orbit bunches grows with a significantly
 lower rate than the number of all orbits.}
\label{fig8}
\end{figure}

Since in classically chaotic systems the number of periodic orbits grows
exponentially with the orbital period or action, a {\em linear} increase of 
the length $s_{\max}$ of the periodic orbit signal \eqref{propagator} requires 
an {\em exponential} increase of the classical input data and the numerical 
effort for the periodic-orbit search.
For the hydrogen atom in a magnetic field at scaled energy $\tilde E=0.5$
the exponential increase of the number of trajectories with growing action
is illustrated by the dashed green line in Fig.~\ref{fig8}.
For the construction of the periodic-orbit signal \eqref{propagator} with
signal length $s_{\max}/2\pi=17$ more than one million periodic orbits must 
be computed.

To reduce the required amount of periodic-orbit data we take advantage of 
properties of the periodic-orbit bunches.
Since all orbits of a bunch are near action-degenerate (cf.\ Fig.~\ref{fig6})
and also have very similar stability properties
and identical Maslov indices the parameters $\mathcal{A}_{\text{po}}$ 
and $s_{\text{po}}$ in Eq.~\eqref{propagator} are approximately the same 
for all members of a periodic-orbit bunch.
(Additional weight factors which arise from symmetry decomposition 
will be discussed in Sec.~\ref{sec:sym_decomp}.)
The idea now is not to compute all members but only one representative
(or very few representatives) of a periodic-orbit bunch, and to use the 
actions and appropriately weighted amplitudes of the representatives for 
the construction of the periodic-orbit signal.
The Maslov index which determines the complex phase of the amplitudes 
in Eq.~\eqref{loading} can be obtained directly and without any numerical 
periodic orbit search from the symbolic dynamics of orbits and reads
$\mu_{\text{po}} = 3L - N_+ - N_-$, where $L$ is the code length and $N_\pm$ 
are the numbers of plus and minus symbols in the code.
Note that the Maslov index does not change by application of any of the
four reconnection rules.

The construction of the reduced data set is achieved as follows.
We need the symbolic dynamics of all orbits and the periodic-orbit parameters
of representatives up to a certain action $s_{\max}$.
In our computations we use $s_{\max}/2\pi=20$.
In a first step we calculate the symbolic sequences of orbits.
As some orbits with very long symbolic dynamics can contribute to 
the periodic-orbit signal the computing time is reduced by optimizing 
the order of generating the symbolic codes in a way that orbits with 
short action are preferably obtained earlier than orbits with long action.
The length limit of every symbolic sequence describing one periodic orbit is 
achieved if the approximate estimate of the action based upon the symbolic 
sequence of every orbit exceeds the action limit $s_{\max}$.
From the symbolic code of a periodic orbit its action can roughly be estimated
by weighting the actions of the fundamental orbits \texttt{0}, \texttt{+}, 
and \texttt{-} with the corresponding number of symbols in the code.
We can calculate efficiently up to a maximum code length $L=16$, i.e., 
the orbit data set is complete up to this length.
Longer symbolic sequences are generated by adding minus symbols to 
shorter symbolic sequences.
This does not change the shape of the orbits or the size of the groups in 
the most cases but adds only new loop parts in outward direction to the 
periodic orbits.
In order to decrease computing time the same method is used to calculate
all periodic orbits.

We use the reconnection rules to group the periodic orbits.
Arranging the orbits is a much faster procedure than the calculation of 
the actions because it is based only on integer arithmetic, which is very 
fast to handle on the computer.
As the parameters of the periodic orbits in one group are very similar,
we can choose one orbit as representative of the group and calculate 
all parameters of this trajectory.
The results with appropriate weighting are then used for the construction 
of the periodic-orbit signal.

The number of periodic-orbit bunches with action $s_{\text{po}}<s$ is shown as
the dotted blue line in Fig.~\ref{fig8}.
Similar to the total number of orbits the number of the bunches also grows
exponentially with the action, but with a significantly lower rate.
For example at $s/2\pi=17$ there are only about $80\,000$ bunches, compared 
to more than one million individual periodic orbits.

\subsection{Symmetry decomposition}
\label{sec:sym_decomp}
The diamagnetic hydrogen atom possesses a discrete symmetry, viz.\ the 
reflection at the $z=0$ plane, and thus the quantum system has resonances
$w_k^\pm$ in the decomposed subspaces with even and odd $z$ parity.
In periodic-orbit theory the symmetry decomposition is achieved by
multiplying the amplitudes $\mathcal{A}_{\text{po}}$ in Eq.~\eqref{loading}
with weight factors $\sigma_{\text{po}}$, which depend on the chosen subspace 
and symmetry properties of the periodic orbits \cite{Cvi93}.
The weight factors are
\begin{align}
 \sigma_{\text{po}} = \left\{
 \begin{array}{ll}
  1 &\text{ for even parity,} \\[2pt]
  (-1)^{N_+} &\text{ for odd parity,}
 \end{array} \right.
\end{align}
with $N_+$ the number of plus symbols in the symbolic code.
In spectra with even parity all members of a periodic-orbit bunch have 
identical weight factors $\sigma_{\text{po}}=1$ and contribute with 
approximately the same amplitudes $\mathcal{A}_{\text{po}}$ to the 
semiclassical signal in Eq.~\eqref{propagator}, i.e., all orbits of a 
periodic-orbit bunch interfere constructively and the total weight of
the bunch amplitude is approximately the amplitude of a representative 
multiplied with the number of orbits in the bunch.
In odd parity spectra the weight factors $\sigma_{\text{po}}$ of two orbits 
which are related by a single application of the reconnection rule R-1 
have different sign and thus the contributions of the two orbits to 
the semiclassical signal in Eq.~\eqref{propagator} approximately cancel.
The total weight of the bunch is the nontrivial sum of the $\sigma_{\text{po}}$ 
of all individual orbits of the bunch, however, that weight can be obtained 
solely from the symbolic dynamics of orbits.

\subsection{Results and discussion}
\label{subsec:results}
For the hydrogen atom in a magnetic field exact quantum and semiclassical
resonances have already been compared by Tanner et al.\ \cite{Tan96}.
The quantum resonances were obtained by numerical diagonalization of the 
generalized eigenvalue problem \eqref{eq:gen_ev} using a complex rotated 
complete basis set.
The semiclassical resonances were computed with a modified and extended 
cycle expansion technique.
As already mentioned at sufficiently high energies $\tilde E>\tilde E_c=0.329$ 
the symbolic dynamics is complete without any pruning of orbits, however, 
trajectories going far away from the nucleus in the direction of the magnetic 
field axis were found to be marginally stable.
Furthermore, resonances accumulate at the thresholds of the Landau channels.
For these reasons the periodic orbit quantization of the diamagnetic hydrogen
atom is a very nontrivial and challenging task.

\begin{figure}
\includegraphics[width=0.88\columnwidth]{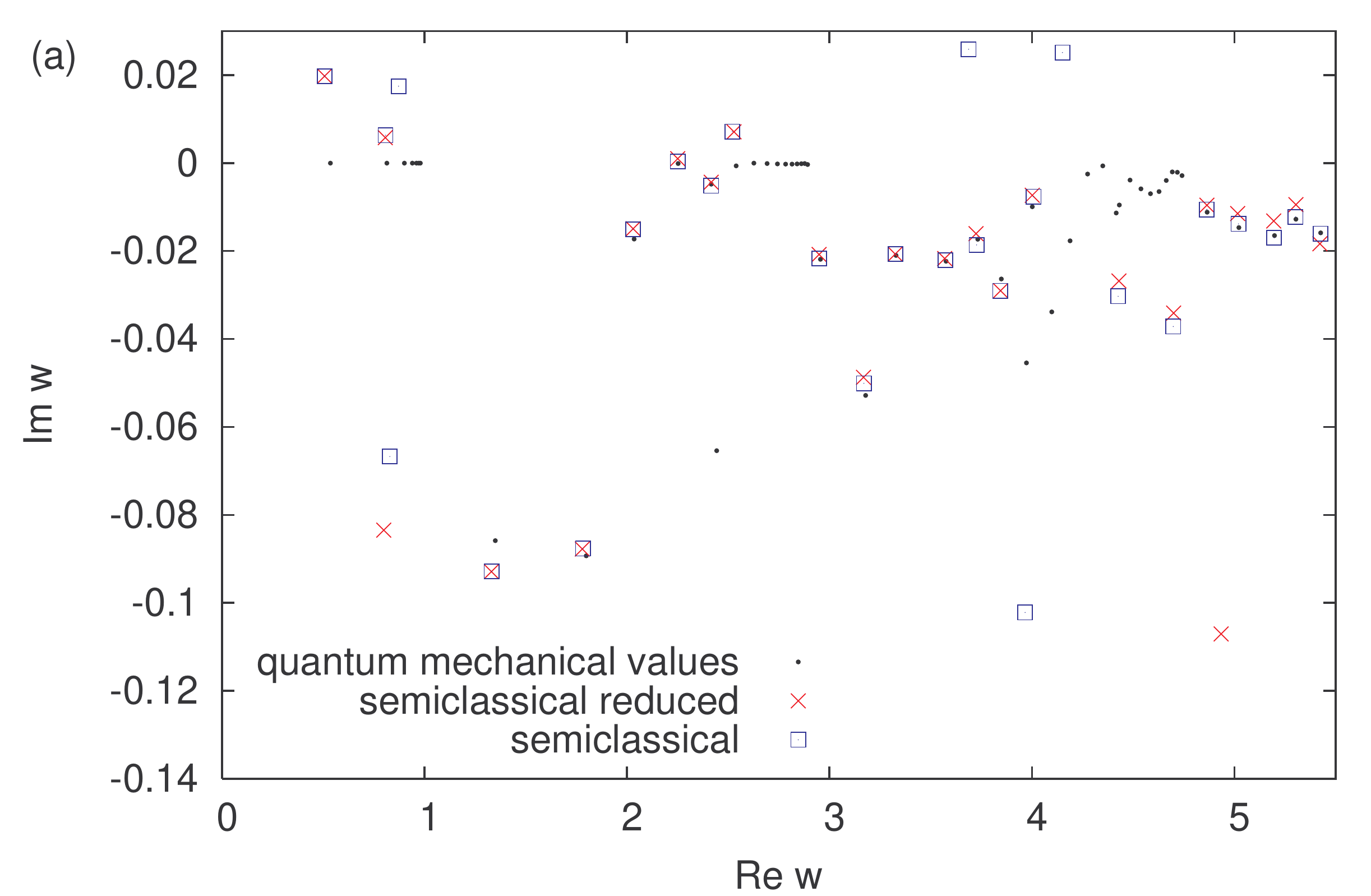}
\includegraphics[width=0.88\columnwidth]{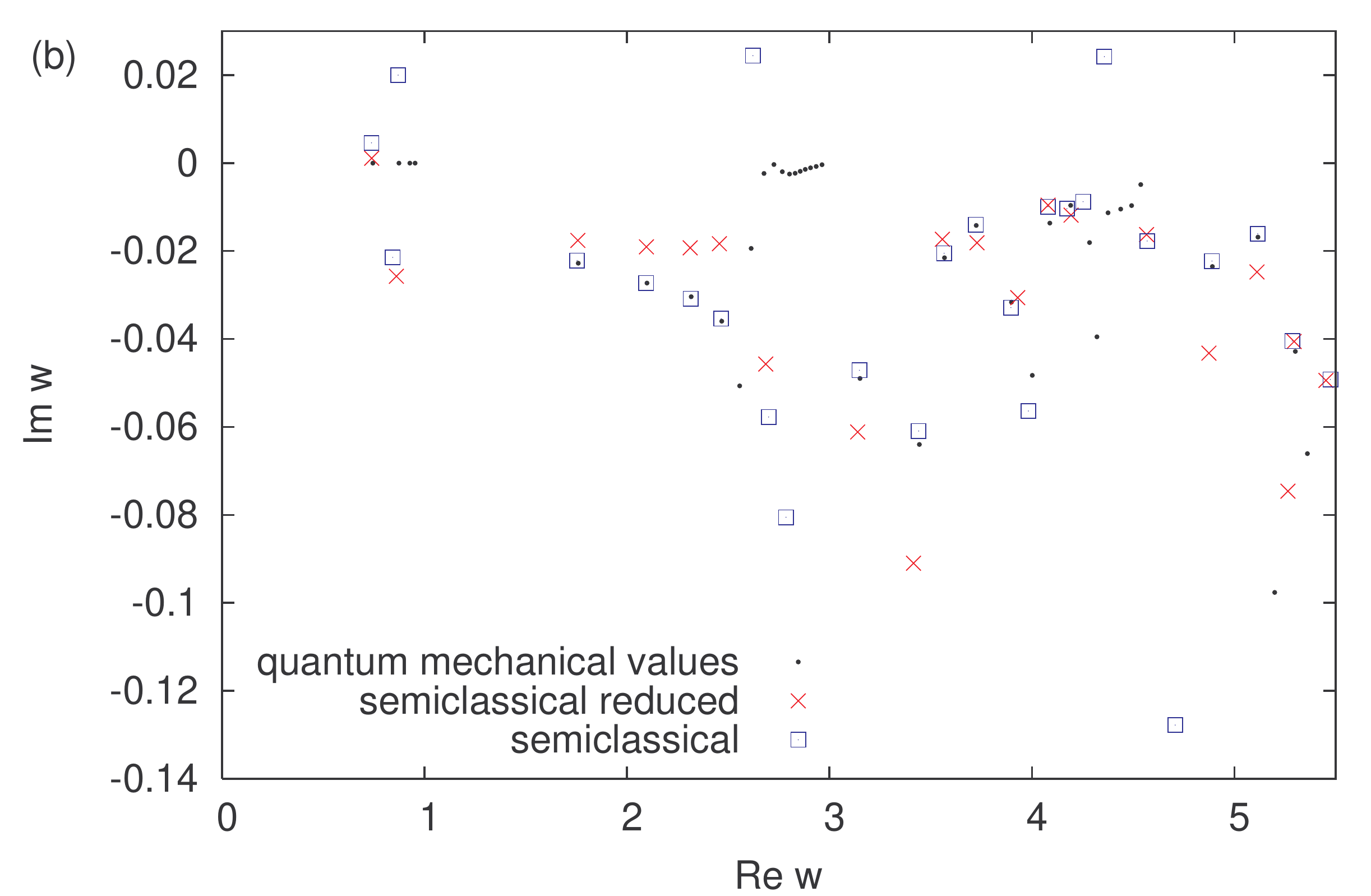}
\caption{(Color online)
 Resonances with (a) even and (b) odd parity of the diamagnetic hydrogen atom
 at scaled energy $\tilde E=0.5$.  In most cases especially the real parts
 of the semiclassical resonances obtained with the reduced data set (red
 crosses) agree very well with the semiclassical resonances obtained with
 the complete periodic-orbit set (blue squares).  For completeness and
 comparison the numerically exact quantum resonances are shown as black dots.}
\label{fig9}
\end{figure}
Here we apply the harmonic inversion method introduced in Sec.~\ref{subsec:harm}
for period-orbit quantization.
With a finite length signal this method can resolve sufficiently isolated
resonances in the spectra but not the accumulation of resonances at the 
Landau thresholds.
However, we do not focus on the comparison of the quantum
and semiclassical resonances but on the results of the two semiclassical
quantizations obtained with either the complete periodic-orbit set or 
with the reduced data set of the periodic-orbit bunches introduced 
in Sec.~\ref{subsec:dataset}.

Resonances with even and odd parity of the diamagnetic hydrogen atom at 
scaled energy $\tilde E=0.5$ are presented in Fig.~\ref{fig9}.
The semiclassical resonances have been obtained by harmonic inversion
of the periodic-orbit signal $C_{\text{sc}}(s)$ with $s_{\max}/2\pi=20$
using either the complete periodic-orbit set (blue squares in 
Fig.~\ref{fig9}) or the reduced data set (red crosses).
In the reduced data set we have used the two representatives with maximum 
or minimum action of each periodic-orbit bunch, which can be identified 
by their symbolic codes.
The numerically exact quantum resonances are shown by black dots in
Fig.~\ref{fig9}.

As mentioned above the harmonic inversion of a finite length signal cannot
reproduce resonances close to the accumulation points at, e.g., $w=1$ and 
$w=3$.
For most of the semiclassical resonances with even parity in Fig.~\ref{fig9}(a)
the agreement is excellent for both the real and imaginary parts of the 
resonances.
For the resonances with odd parity in Fig.~\ref{fig9}(b) the agreement
of the imaginary parts is less perfect, however, the real parts still agree
very well.
The reason for the larger deviations in spectra with odd parity might be
that the possible cancellation of periodic orbit contributions in this 
subspace is more critical for the accuracy of the periodic-orbit signal 
than the solely constructive superposition of the contributions in the 
subspace with even parity.
The results clearly demonstrate that the use of near action-degenerate
periodic-orbit bunches allows for the reduction of the classical data set
by more than an order of magnitude and thus significantly increases the
efficiency of periodic-orbit quantization methods.

\section{Conclusion}
\label{sec:conclusion}
The existence of periodic-orbit bunches in the classically chaotic diamagnetic
hydrogen atom has been revealed.
The orbits of a bunch have a similar shape in the fundamental domain of 
the coordinate space and only differ in the behavior at self-encounters.
We have introduced four reconnection rules which allow for the grouping of
orbits already on the level of the symbolic dynamics in terms of a ternary
alphabet.

The exponential proliferation of periodic orbits of a classically chaotic
system implies that semiclassical methods based on Gutzwiller's trace
formula typically require an exponentially growing classical data set
to resolve more eigenstates.
We have shown that the use of one or few representatives of the near 
action-degenerate orbits can help to significantly improve the efficiency
of semiclassical quantization methods.

A peculiarity of the diamagnetic hydrogen atom is that periodic orbits
at positive energies are marginally stable.
It will thus be very interesting to investigate the existence of near 
action-degenerate periodic orbits for other chaotic systems such as
$N$-disk or $N$-sphere scatterer, and to study the improvement
of semiclassical quantization methods by using the periodic-orbit bunches.

%

\end{document}